\documentclass[prb,twocolumn]{revtex4}

\usepackage{latexsym,amsmath,graphics,graphicx}
\usepackage [english] {babel}

\begin{document}
\title{Cd-Vacancy and Cd-Interstitial Complexes in Si and Ge}
\author{H.~H\"ohler, N.~Atodiresei, K.~Schroeder, R.~Zeller, and P.~H.~Dederichs}
\affiliation{Institut f\"ur
Festk\"orperforschung, Forschungszentrum J\"ulich, D-52425 J\"ulich,
Germany}
\date{\today}

\begin{abstract}
  The electrical field gradient (EFG), measured e.g. in perturbed
  angular correlation (PAC) experiments, gives particularly useful
  information about the interaction of probe atoms
  like$\phantom{x}^{111}\mathrm{In}/\phantom{x}^{111}\mathrm{Cd}$ with
  other defects. The interpretation of the EFG is, however, a
  difficult task. This paper aims at understanding the interaction of
  Cd impurities with vacancies and interstitials in Si and Ge, which
  represents a controversial issue. We apply two complementary {\it ab
    initio} methods in the framework of density functional theory
  (DFT), (i) the all electron Korringa-Kohn-Rostoker (KKR)
  Greenfunction method and (ii) the Pseudopotential-Plane-Wave (PPW)
  method, to search for the correct local geometry. Surprisingly we
  find that both in Si and Ge the substitutional Cd-vacancy complex is
  unstable and relaxes to a split-vacancy complex with the Cd on the
  bond-center site. This complex has a very small EFG, allowing a
  unique assignment of the small measured EFGs of 54MHz in Ge and
  28MHz in Si. Also, for the Cd-selfinterstitial complex we obtain a
  highly symmetrical split configuration with large EFGs, being in
  reasonable agreement with experiments.
\end{abstract}


\maketitle

\section{\bf Introduction}
The detailed understanding of the electronic structure of intrinsic defects
 in semiconductors is in many cases still an unresolved problem.
 Many different experimental methods aiming at microscopic information about
 intrinsic defect structures exist. 
The ``state of the art'' experiments for these kind of 
problems are the measurements of hyperfine interactions. One example
is the perturbed angular correlation measurement (PAC) of the electric field
gradient (EFG) \cite{Wichert:89.1,Wichert:89.2,Wichert:99.1,Forkel:92.1,Acht:93.1}.
The EFG describes the interaction of the electrons with the electric quadrupol
moment $Q_{ij}$ of the nuclei and is given by the second derivative $V_{ij}$ of the
Coulomb potentials at the nuclear site.

Since the PAC technique itself is based on a radioactive decay, it is essential for Si and Ge host crystals to implant a proper probe atom. The transformation  of implanted $\phantom{x}^{111}\mathrm{In}$ isotopes to$\phantom{x}^{111}\mathrm{Cd}$ by electron capture leaves the$\phantom{x}^{111}\mathrm{Cd}$ probe in an excited state and the subsequent decay to its ground state makes the determination of the EFG possible. That means the EFG is just measured at the Cd site and reflects the local electronic structure which again depends on the atomic environment.

Since the EFG vanishes for cubic or tetrahedral point symmetry, probe atoms like
Cd give no signal on substitutional sites. An EFG only arises due to the symmetry
lowering induced by other defects. In addition the direction of the main axis of the
EFG gives very clear information about the symmetry of the corresponding complexes
formed. Therefore PAC measurements yield particularly useful information about the
interaction of Cd probe atoms with other defects. The only disadvantage of the method
is that the resulting EFGs are difficult to understand and that reliable 
information about the EFG and the electronic and geometrical structure of complexes can
only be obtained by ab-initio calculations.

The paper aims at understanding PAC experiments with Cd probes in Si\cite{Forkel:87.1,Deicher:87.1,Wichert:89.3} and Ge\cite{Siele:98.2,Feuser:90.1,Siele:98.1,Siele:97.1}. Of particular interest is the interaction with intrinsic defects, i.e. vacancies and selfinterstitials. In previous studies two complexes in Ge have been identified, both with main axis along (111) and asymmetry parameter $\eta=0$ but with very different quadrupolar coupling constants $\nu$, i.e. $\nu_1=54\mbox{MHz}$ and $\nu_2=415\mbox{MHz}$. Both values, which differ by an order of magnitude, were measured\cite{Siele:98.2,Feuser:90.1} by two different groups and interpreted as Cd complexes with intrinsic defects. However the two groups disagree about the assignment of the EFG. For Feuser et al. $\nu_2$ is assigned to a substitutional Cd-vacancy complex, while $\nu_1$ should refer to an interstitial Cd atom in the vicinity of a vacancy. In contrast to this Sielemann et al.\cite{Siele:98.1} assign $\nu_1$ to the substitutional Cd atom 
with a vacancy on the nearest-neighbor site and $\nu_2$ to a substitutional Cd
with a selfinterstitial on a (111) oriented interstitial position. Compared to Ge, the interpretation of the measured frequencies is much more uncertain in Si\cite{Wichert:89.3}.

In this paper we perform ab-initio calculations to solve this problem. Basically we can confirm the assignment of Sielemann et. al\cite{Siele:98.2,Siele:98.1,Siele:97.1} that the small frequency $\nu_1$ belongs to the Cd-vacancy complex and the high frequency to the interstitial one. However in both cases the configurations are very unusual: The Cd atom sits on a bond-center position and the single vacancy as well as the interstitial are split on two neighboring sites. The existence of such an impurity-split vacancy configuration has already been suggested by Watkins\cite{Watkins:75.1} for a Sn-vacancy pair in Si. In a forth-coming paper we will show that this pair configuration is also the most stable one for the vacancy complexes with other impurities like In, Sn and Sb in Si and Ge. Therefore we expect that our results are typical for oversized impurities in group IV semiconductors. 
 
\section{Theoretical Methods}
All calculations are based on density functional theory in the local density
approximation\cite{Vosko:80.1}. Two different methods have been used to solve the Kohn-Sham equations.
\begin{table*}[t]
\begin{center}
\begin{tabular}{|c|c|c|c|c|c|c|}
\hline
atom & pp-type & atom conf. & s-channel & p-channel & d-channel & f-channel \\
\hline
Ge & KB & ground state & 1.15 & 1.15 & 1.65 & - \\
\hline
Si & KB & ground state & 1.15 & 1.25 & 1.25 & - \\
\hline
Cd & PAW & ground state & 2.33 / -0.0408 & 2.6 / -0.1052 & 2.6 / -0.105 & -   \\

       & & ground state & 2.33 / -0.1 & 2.6 / 0.4 & 2.6 / -0.6 & 2.6 / -0.4 \\
\hline
\end{tabular}
\caption{Parameters used to construct the pseudopotential projectors. 
For the KB-potentials only the cut-off radii are listed for each l-channel, 
for the PAW-potentials the cut-off radii / reference energies (in Ry) are listed. 
The first line for Cd refers to the eigenvalues of the bound states for the
respective l-channel, the second line to the specific energies for the unbound states
used to construct the second projector for each l-channel. All radii are in $a_B$
(0.529177 \AA). 
\label{table:gen_proj}}   
\end{center}
\end{table*}
The first one is the pseudopotential plane wave (PPW) method, which has mostly been 
used to investigate the configuration and stability of the Cd-split vacancy and the
Cd-split interstitial configurations. These configurations, as well as the relaxations
of the neighboring atoms, have been recalculated by the KKR-Greenfunction method\cite{Korr:47.1,Kohn:54.1,Brasp:84.1}, which as an all electron method allows also to calculate electric field gradients and hyperfine fields\cite{Akai:90.1,Blaha:88.1,Blaha:88.2}.

The PPW method approximates the inhomogeneous system containing defect 
complexes by periodically arranged large supercells
and uses plane waves to expand the electronic wave functions. 
This has the advantage that band-structure methods can be used to determine 
the electronic structure, and total energies and forces on the atoms can be calculated
without difficulty for arbitrary arrangements of the atoms in the supercell.
Our {\tt{EStCoMPP}}-program\cite{kromen:01} can handle norm-conserving Kleinman-Bylander
(KB)-pseudopotentials\cite{K-B:82} and projector-augmented 
(PAW)-multi-projector-pseudopotentials\cite{Bloechl:94} used to describe the electron-ion-interaction. 
For Si and Ge KB-non-local pseudopotentials are used for the $s,p$-valence electrons 
and the $d(l=2)$-component is used as a local potential. 
For Cd the $5s,p$-valence electrons as well as the the $4d$-electrons are 
treated explicitly with PAW-potentials and the $f(l=3)$ -component is used 
as a local potential. Partial-core-correction was used for Cd. 
The atomic configurations, cut-off radii and reference energies used to generate the
projectors are listed in Table~\ref{table:gen_proj}.
For the actual calculations we used a (111)-oriented supercell with the basis vectors
$\vec{b}_1 = 3 a (0, -1, 1); \vec{b}_2 = 3 a (-1, 1, 0); \vec{b}_3 = 2 a (1, 1, 1)$
containing 108 atoms. $a$ is the theoretical lattice constant ( 10.44 $a_B$ for Ge
and 10.25 $a_B$ for Si; Bohr radius $a_B$ = 0.529177 \AA). 
The Cd-complex was placed in the middle of the cell. 
We used $C_{3v}$ symmetry explicitly for all configurations.
A plane-wave basis set equivalent to $6\times6\times6$ 
Monkhost-Pack\cite{Monkhort-Pack:77} $\vec{k}$-points was used
which yields three inequivalent $\vec{k}$-points in the irreducible part of the 
Brillouin-zone for our supercell. We used a plane-wave cut-off of 20.25 Ry, which is
required for the convergence of the Cd PAW-potential. 
The atoms belonging to the Cd-complexes and their nearest neighbors were relaxed 
until the forces on all atoms were less than 0.1 mRy/$a_B$. 
We checked that forces on further neighbors (which were not moved) were of the order of
1 mRy/$a_B$ and we estimate that the positions of all relaxed atoms were determined with
an accuracy better than 0.01 $a_B$.

In the KKR Greenfunction method the calculation is devided in two steps. First the Greenfunction of the host is determined. In a second step the host Greenfunction is used to determine the Greenfunction of the crystal with a single impurity by a Dyson equation. For details we refer to reference\cite{Nikos:97.1}. All calculations are performed with an angular momentum cut-off of $l_{max}=4$. For the host crystal we use the LDA lattice constants\cite{Settels:99.3} of Si (10.21$a_B$) and Ge (10.53$a_B$). A k-mesh of $30\times 30\times 30$ $\vec{k}$-points in the full Brillouin zone is used. We use the screened KKR-formalism\cite{Zeller:95.1,Zeller:97.1}, with the TB-structure constants determined from a cluster of 65 repulsive potentials of 4 Ry heights. The diamond structure is described by a unit cell with 4 basis sites, two for host atoms and two for vacant sites. For the Greenfunction of the defective systems, we allow 77 potentials of the defect and the surrounding host atoms to be perturbed, which are then calculated selfconsistently with proper embedding into the host crystal. All calculations include the fully anistropic potentials in each cell and thus allow the reliable calculation of forces, lattice relaxations and electric field gradients. The single particle electron density $n(\vec{r})$ can be calculated via the imaginary part of the Greenfunction:
\begin{eqnarray}
\label{ng}
n(\vec{r})=-\frac{2}{\pi}\Im{m}\int_{-\infty}^{E_f}dE\rule{0.2cm}{0cm}{G(\vec{r},\vec{r};E)}.
\end{eqnarray}
The  Coulomb potential can be expressed in terms of $n(\vec{r})$ and therefore the force $\vec{F}^n$ on atom $n$ can be derived with the use of the ``ionic'' Hellmann-Feynman theorem\cite{Nikos:97.1}:
\begin{eqnarray}
\vec{F}^n= & Z^n \frac{\delta V_{coul}}{\delta\vec{r}}\vert_{\vec{r}=\vec{R}^n}
-\int d\vec{r}n_C(\vec{r}-\vec{R}^n)\frac{\delta V_{eff}}{\delta\vec{r}},
\end{eqnarray}
where the first term denotes the Coulomb force of the valence electrons and of the other nuclei on the nuclear charge $Z^n$, while the second term gives the force on the core electrons (C) of atom n, evaluated by the derivative of the Kohn-Sham potential. It can be shown, that in an angular momentum expansion the force is determined via the $l=1$ component of the charge density. The tensor of the EFG is given by the second derivative of the Coulomb potential: 
\begin{eqnarray}
\label{defefg}
{\bf V}_{\alpha\beta}= & \Biggr\{\frac{\partial V_{coul}}{\partial{\alpha}\partial{\beta}}-\delta_{\alpha\beta}\frac{1}{3}\sum_{i=1}^3\frac{\partial V_{coul}}{\partial{\alpha}^2}\Biggl\}\Biggl\vert_{\vec{r}=0} ,\\ \nonumber & \alpha,\beta=x,y,z;\hspace{0.25cm}\mbox{Tr }{\bf V}_{\alpha\beta}=0,
\end{eqnarray}
where $(x,y,z)$ denotes the coordinate system. In practical applications, since we expand all quantities in spherical harmonics, the electric field gradient is calculated as follows\cite{Blaha:88.1,Blaha:88.2}:
\begin{eqnarray}
\label{defefg2}
{\bf V}_{\alpha\beta}  = & \sum_{m=-2}^2 \tilde{V}_{2,m}(0)\partial_{\alpha}\partial_{\beta}(r^2Y_{2m}(\vec{r})),\\
\nonumber & \tilde{V}_{2,m}(0) = I_1 \ + \ I_2,\\
\nonumber & I_1 =  \frac{8\pi}{5}\int_0^{R_{MT}} {r'}^2 \frac{n_{2m}(r')}{{r'}^3}dr'; \\
\nonumber & I_2 =  \frac{V_{2m}(R_{MT})}{{R_{MT}}^2}-\frac{8\pi}{5}\int_0^{R_{MT}} \frac{n_{2m}(r'){r'}^4}{{R_{MT}}^5}dr',
\end{eqnarray} 
where $Y_{2m}(\vec{r})$ are spherical harmonics, $\tilde{V}_{2,m}(r)$ are the $l=2$ components of the Coulomb potential and $R_{MT}$ is the Muffin Tin Radius of the probe atom. Furthermore $n_{2m}(r)$ are the $l=2$ components of the radial charge density. We denote that just the $l=2$ components are needed for the EFGs, while the $l=1$ components enter into the forces. Due to this, the EFG vanishes for sites with cubic or tetrahedral symmetry. For the cell division of the crystal we used a generalized Voronoi construction. The impurity cell at the bond center was constructed such that it is slightly larger than the cells of the host atoms, thus avoiding to decrease the Muffin Tin Radius of real atoms by more than 16\%. For the energy contour in equation (\ref{ng}) all $d$-states are included. For the treatment of occupied states in the band gap, we have used a separate energy contour for gap energies. Since we used group theory we are able to occupy these gap-states according to their symmetry, in this way avoiding to calculate the wave functions of the localized gap states.  
\section{Cd-SPLIT VACANCY COMPLEX}
\begin{figure}[t]
\begin{center}
\includegraphics[scale=0.6]{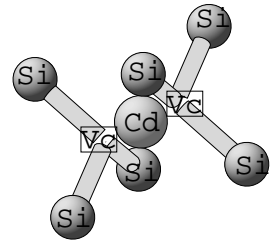}
\caption{The Cd-split vacancy configuration}
\label{fig:1}
\end{center}
\end{figure}

{\em The substitutional Cd-vacancy complex:} The starting point of our study is the substitutional Cd-vacancy complex. To our surprise this complex with the Cd on the substitutional site and the vacancy on the nearest neighbor site is highly unstable. When fixing the Cd at the substitutional position and allowing the six host atoms to relax, the energy of the configuration is for both Si and Ge hosts more than 1 eV higher than the energy of the stable split configuration shown in Fig.\ref{fig:1}. When the Cd atom is allowed to relax, the energy continuously lowers (without energy barrier) until the Cd reaches the stable bond-center position shown in Fig.\ref{fig:1}. The fixed substitutional configuration is highly anisotropic and the calculations give rather high EFGs for the Cd atom, with values of about -420 to -430 MHz in Ge and -390 to -420 MHz in Si, where the spread is caused by the different charge states. However both the PPW and the KKR calculations show that this configuration is unstable and does not represent a local minimum. Thus a comparison of the EFGs with the measured frequency $\nu_2 = 415$ MHz in Ge is meaningless, since the configuration is not stable. Apparently the instability arises from the large size of the Cd atom, which is pushed away from the three Si or Ge neighbors into the empty space towards the vacancy and settles on the symmetrical bond-center position.
\begin{figure}[b]
\begin{center}
\includegraphics[scale=0.41]{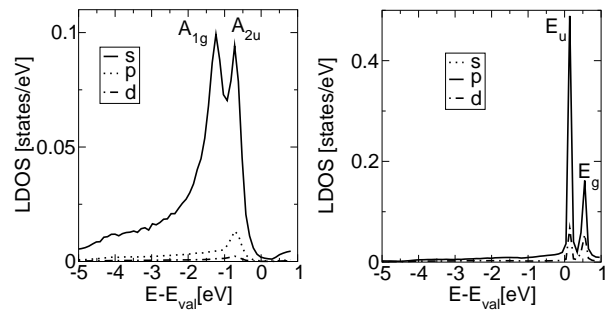}
\caption{Local density of states (LDOS) for the $\mbox{divacancy}^{2+}$ in Si. Shown are the local s, p and d contributions on a vacancy site, separately for the $A_{1g}$ and $A_{2u}$ subspaces(left) and the $E_u$ and $E_g$ subspaces(right). As can be seen, for the A subspace we find mainly s-contributions and for the E subspace mainly p-contributions.
The other l-channels contribute less than 10\% over the entire energy
range.}
\label{fig:2}
\end{center}
\end{figure} 

\begin{figure}[t]
\begin{center}
\includegraphics[scale=0.5]{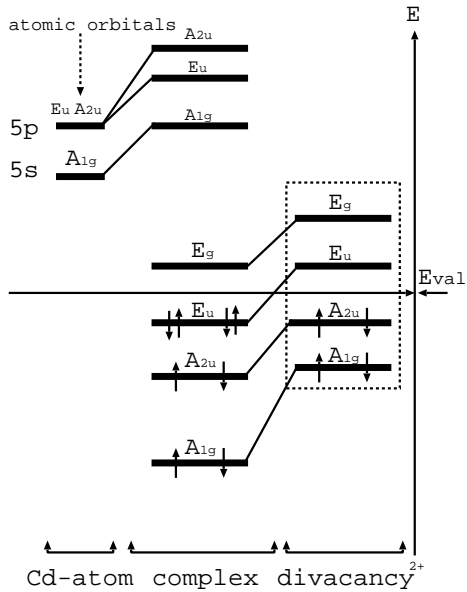}
\caption{Schematic tight-binding scheme of energy levels to explain the
electronic structure of the Cd-split vacancy complex in Si and Ge.
Included are the 5s, 5p levels of the Cd atom and the dangling bond
states of the
divancy in its 2+ state.
From left to right: the single Cd-atom, the Cd-split vacancy complex and the $\mbox{divacancy}^{2+}$}.
\label{fig:3}
\end{center}
\end{figure}

{\em The divacancy in Si and Ge:} Since the electronic structure of the Cd-split vacancy complex can most easily be understood by first considering the divacancy, we present in Fig.\ref{fig:2} the calculated density of states (DOS) of the divacancy in Si. Four different peaks can be seen: single degenerate A$_{1g}$ and A$_{2u}$ resonances in the valence band, and double degenerate $E_g$ and $E_u$ localized states in the gap. The splitting into $g$ and $u$ state arises from the hybridization of the A- and 
E-dangling bond states of the two single vacancies into bonding and anti-bonding combinations. In a molecular orbital (MO) scheme the energy levels are schematically indicated in Fig.\ref{fig:3} on the right. In the calculation the Fermi level is situated at the maximum of the valence band $E_{val}$. Therefore only 4 of the 6 dangling bond electrons can be accommodated, so that in the calculation the charge state of the divacancy is $2+$. 

\begin{figure}[t]
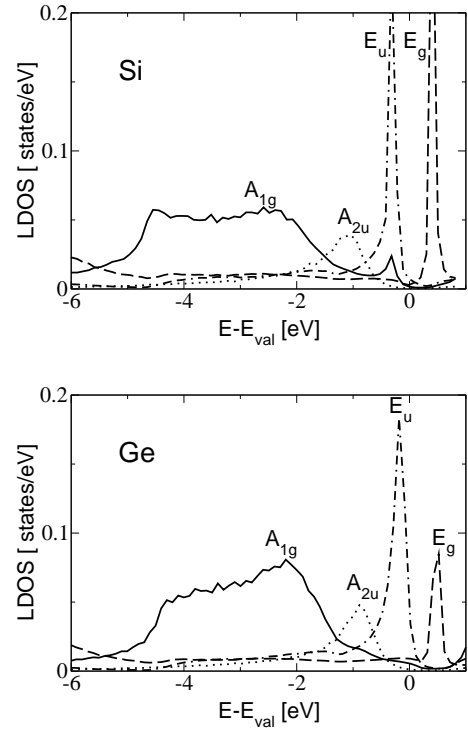

\center{\includegraphics*[scale=0.25]{cd_in_si_sym_paper.eps}\\[0.5cm]
\includegraphics*[scale=0.25]{cd_in_ge_sym_paper.eps}
\caption{LDOS projected on the irreducible subspaces of the $D_{3d}$ group at the Cd site (bc) for the Cd-split vacancy in Si (above) and Ge (below). The nearest neighbors
are fixed on ideal lattice positions.\label{fig:4}}}
\end{figure}

{\em Cd-split vacancy complex:} The local DOS at the Cd bond-center site of the Cd-split vacancy complex is shown in Fig.\ref{fig:4}, both for Si and Ge. Since this complex preserves the point symmetry $D_{3d}$ of the divacancy and since the Cd $5s$ and $5p$ states are well above the Fermi level, the decomposition and level structure of the divacancy DOS are basically preserved. As indicated by the MO level scheme in the center of Fig.\ref{fig:3} , by hybridization between the states of the Cd and of the divacancy the divacancy levels are pushed to lower energies and the Cd states to higher energies. The selfconsistent calculations yield the $E_u$ state as a resonance at the top of the valence band, in agreement with the level scheme assumed for the split complex in Fig.\ref{fig:3}. In this configuration 8 electrons are accommodated in the single degenerate $A_{1g}$ and $A_{2u}$ states and the double degenerate $E_u$ state. These 8 electrons just correspond to the 6 electrons of the dangling bonds of the divacancy and the two electrons of Cd, so that the calculations yields the neutral charge state, as long as the $E_g$ level in the gap remains unoccupied. The $d$-states of Cd are located around $-8.5$ eV and, while included in the calculations, are basically localized and of minor importance for the electronic structure and the EFG. 

As Fig.{\ref{fig:1}} shows, the Cd atom has in its nearest coordination shell six host atoms at a distance at 1.26 times the nearest neighbor distance (if lattice relaxations are neglected; see below). Therefore the hybridization between the Cd $5sp$ states and the divacancy states is rather weak, so that the occupied $A_{1g}$, $A_{2u}$ and $E_u$ states contain locally only relatively small admixtures of these Cd states, leading to a charge of only 1.0 electrons (in the case of Ge; for Si: 0.9 electrons) in the Cd sphere, which has about the same size as the Si or Ge spheres. Therefore we can understand the electronic structure in a simple picture in real space, by considering the Cd as a ''naked'' Cd$^{2+}$ ion embedded at the bond-center position into the divacancy. The strongly attractive Cd$^{2+}$ potential lowers all divacancy states, but preferentially, and in first order perturbation theory only, the $A_{1g}$ level which has the full $D_{3d}$ symmetry of the defect. Due to this, the $A_{1g}$ state is stronger localized at the Cd site than the other states and shows up more prominently in the LDOS of Cd. The LDOS for the Si and Ge hosts are qualitatively and quantitatively very similar. 

The different charge states $[\rm{Cd} \ \rm{V}]^{1-}$, $[\rm{Cd} \ \rm{V}]^{2-}$, $\dots$ of the Cd-split vacancy complex are obtained by occupying the double degenerate $E_g$ state in the gap. The $[\rm{Cd} \ \rm{V}]^{1-}$ state is magnetic and exhibits a small moment about 0.02 $\mu_B$ at the Cd site, both in Si and Ge. Since the total moment is 1 $\mu_B$, this shows the large spatial extend of the $E_g$ state. 

\begin{figure}[t]
\begin{center}
\includegraphics[scale=0.94]{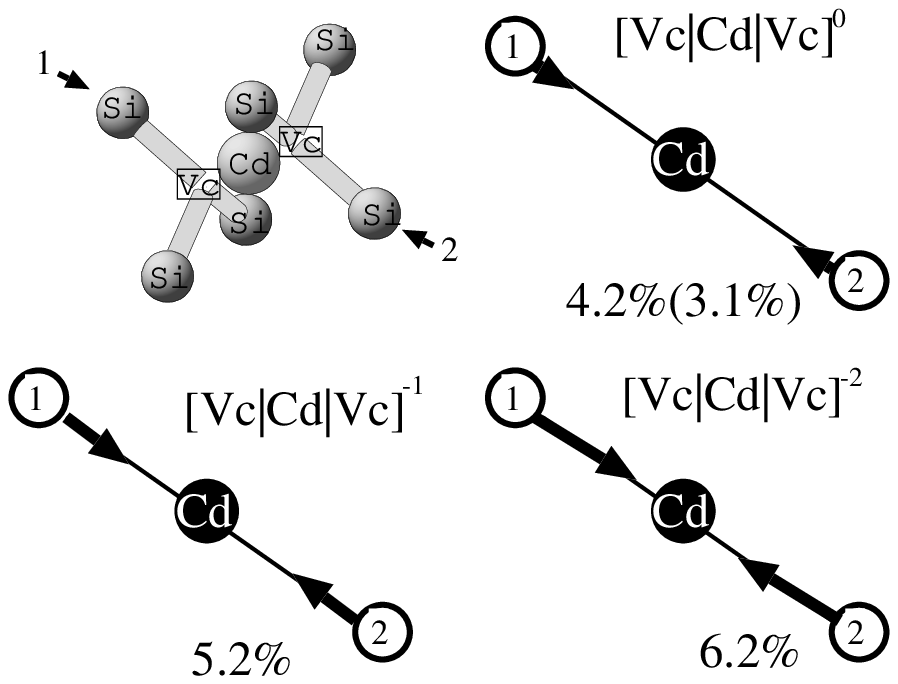}\\[1cm]
\begin{tabular}{|c|c|c|}
\hline
Defect & KKR  [$ \%$ NN] & PPW [$ \%$ NN] \\
\hline
$[V|Cd|V]^0$ in Ge & 6.6 & 6.0 \\
\hline
$[V|Cd|V]^{1-}$ in Ge & 7.5  &  \\
\hline
$[V|Cd|V]^{2-}$ in Ge & 8.5  &  \\
\hline
\end{tabular}
\caption{Relaxations of the 6 nearest neighbors of the Cd-atom, Cd-split vacancy complex in the figure for the Si host, in the table for the Ge host. Besides the KKR values for the neutral configuration also the PPW values are given.\label{fig:5}}   
\end{center}
\end{figure}

{\em Lattice Relaxations:} Fig.\ref{fig:5} shows the calculated lattice relaxations of the six nearest neighbor Si-atoms around the Cd atoms, while the table gives the calculated values for the case of Ge. All relaxations are given in percentages of the nearest neighbor distance of the host crystal. First, we see that in the case of the neutral complexes the KKR-values agree well with the PPW ones. The small differences arise from the different ways used to describe the electron-ion interaction (pseudopotentials for the PPW and all-electron in KKR) and the slightly different lattice constants. They are of the same order of magnitude as typical differences in other instances, e.g.\ for the relaxations of the near surface atoms of Cu surfaces. Secondly, we notice that the relaxations point toward the Cd atom and increase by about 1 $\%$ with increasing total charge of the complex. Both facts are consequences of the electrostatic interaction, i.e. the Cd$^{2+}$ ion attracts the dangling bond electrons of the Si or Ge neighbors and the attraction increases with increasing charge in the extended $E_g$ states. Third, the relaxations in Ge are larger than the relaxations in Si, which is a consequence of the 3 $\%$ larger lattice constant of Ge as compared to Si (theoretical values), so that the local distances are very similar in both hosts. Despite the sizeable relaxations the final distances are still very large, the strength of the hybridization is still weak and the LDOSs are hardly changed by relaxations. 

{\em Charge anisotropy and EFGs:} As stressed in the introduction the EFG is determined by the anisotropy of the charge density at the Cd site, more correctly by the $\ell = 2$ components of the $(\ell m)$-projected charge density $n_{2m} (r)$. Such $\ell = 2$ components arise from the anisotropy of the $p$- and the $d$-charge densities, since mixed $s-d$ or $p-f$ terms are very small. In a good approximation the EFG is therefore determined by the anisotropy of the local $p$ and $d$ charge\cite{Blaha:88.1,Blaha:88.2} 
\begin{equation}
V_{zz} \ = \ a \, \Delta N_p + b \, \Delta N_d
\end{equation}
where $\Delta N_p$ and $\Delta N_d$ are the anisotropic $p$ and $d$ charges, which for the considered symmetry and $z$-axis in (111) direction are given by
\begin{eqnarray}
\Delta N_p =& 
\int^{E_F}_{0} dE' [\frac{1}{2} \ \left( n_{p_x} (E') + n_{p_y} (E') \right) - n_{p_z} (E')] \nonumber \\
\Delta N_d =& 
\int^{E_F}_{0} dE' \ [n_{d_{x^2-y^2}} (E') + n_{d_{xy}} (E') \label{aniso} \\
& \ - \ \frac{1}{2} \ \left( n_{d_{xz}} (E') + n_{d_{yz}} (E') \right) - n_{d_{z^2}} (E')]\nonumber
\end{eqnarray}
The positive constants $a$ and $b$ in equation (\ref{aniso}) are matrix elements of $\frac{1}{r^3}$ with normalized $p$ and $d$ wave functions at $E_F$.

We can now analyze the anisotropic charges in terms of the contributions from the different irreducible contributions $A_{1g}$, $A_{2u}$, $E_u$ and $E_g$ to the local Cd DOS as given in Fig.\ref{fig:4}. The $A_{1g}$ subspace only contains the local $s$ and $d_{z^2}$ contributions, the subspace $A_{2u}$ only the $p_z$, the $E_u$-subspace the $p_x$ and $p_y$ and the $E_g$ state the local $d_{xy}$, $d_{x^2-y^2}$, $d_{xz}$ and $d_{yz}$ contributions. 

In the neutral state $[\rm{CdV}]^0$ the $E_g$ state is not occupied and we find negative electric field gradients. The contribution $\Delta N_p$ is negative, since the $E_u$ state cannot overcome the negative $n_{p_z}$ contribution of $A_{2u}$ and $\Delta N_d$ is negative due to the $n_{d_{z^2}}$ contribution from $A_{1g}$. When one then populates the $E_g$-state, we find that the contributions from $d_{x^2-y^2}$ and $d_{xy}$ are very small, so that additional negative contributions from $d_{xz}$ and $d_{yz}$ are expected for the charged configurations.  

The calculated values for 
the relaxed configurations are given in table \ref{table.1}.
\begin{table}[b]
\centerline{ 
\begin{tabular}{|l|c|c|c|c|}
\hline
host crystal &       Si         & Ge & Si & Ge\\
        &  EFG [MHz]              & EFG [MHz] & HF [kG] & HF [kG] \\
\hline
$[V|Cd|V]^{0}$ & -6.97 & -32.69 & &\\
\hline
$[V|Cd|V]^{1-}$ & -27.99 &  -55.69 & 21.8 & 8.5 \\
\hline
$[V|Cd|V]^{2-}$ & -49.47 &   -79.64 & & \\
\hline
EXP            & $\pm$ 28.00  & $\pm$ 54.00 & & \\
\hline
\end{tabular}}
\caption{Calculated EFG on fully relaxed positions for the Cd-split vacancy in Si and Ge for different charge states ($0,1-,2-$). We used Q=0.83barn for Cd. Experimental values (EXP) are taken from reference \cite{Siele:98.2,Forkel:87.1}. \label{table.1}}
\end{table}
First, we would like to point out that the displacements of the NN-atoms play only a minor role, which is not surprising in view of the small changes found in the LDOS. 
Secondly, we observe that for the charged defects the EFGs in Si and Ge increase for each electron added in the $E_g$ state by about 20 MHz, more correctly about 21 in Si and 23 in Ge. This is a consequence of the fact that the $E_g$ state is very extended: For each electron added to the complex the local charge of Cd increases only by 0.05 electrons, which represents a weak perturbation. Finally we find that the EFG of the singly charged configuration $[CdV]^{1-}$ agrees both in Si and Ge very well with the measured values. Note that experimentally the sign of the EFG cannot be determined. We consider the close agreement as a confirmation that the Cd-split vacancy complex is the most stable Cd-vacancy complex and that the measurements refer to the negatively charged $[\rm{CdV}]^{1-}$ complex. This is in agreement with Haesslein et al.\cite{Siele:98.2}, who assigned the frequency $\nu_1 = \pm 54$ MHz in Ge to a Cd-vacancy complex in the charge state $1-$. As mentioned earlier already, in this configuration the complex is magnetic and for the Cd-atom our calculations predict a small hyperfine field (HF) of 21.8 kG in Si and 8.5 kG in Ge. Up to now these hyperfine fields have not been measured. In summary we find that the small EFGs measured are a typical feature of the Cd-split vacancy configuration, in which the Cd atom is far away from the nearest neighbors so that the hybridization is very small, much smaller than for the strongly anisotropic (but unstable) substitutional configuration, yielding EFGs of around 400 MHz. Partly the geometrical structure also helps, since the Cd atoms and the six neighbors form a distorted octahedron. For an ideal octahedron the EFG would vanish due to symmetry. 
\section{Cd-INTERSTITIAL COMPLEX}

\begin{figure}[b]
\begin{center}
\includegraphics[scale=0.6]{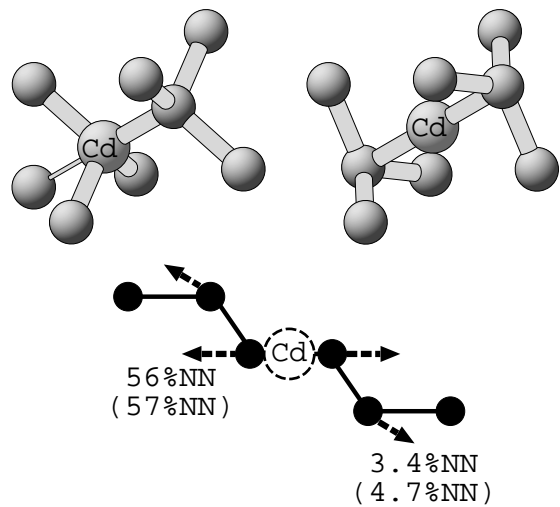}
\caption{The Cd-selfinterstitial complex with Cd on the substitutional site (above left) leads after relaxation to the symmetrical complex with Cd on the bond center and the two host atoms shifted half way between the substitutional and interstitial positions (above right). The lower figure gives the calculated displacements of the neighbors from the ideal positions in percent of the NN distance for the Ge host. In brackets the results of the PPW calculation are given.}
\label{fig:6}
\end{center}
\end{figure}

The Cd-selfinterstitial complex is a possible candidate for the large EFG of $\nu_2 = 415$ MHz measured by two groups\cite{Siele:98.2,Feuser:90.1} in Ge. In our electronic structure calculations we have first studied the interstitial-Cd complex shown in Fig.\ref{fig:6} (upper left): a self-interstitial on a tetrahedral site adjacent to a substitutional Cd impurity. To our surprise we find again, that this configuration is very unstable, exhibiting a large force of about 100 mRyd/$a_B$ on the Cd atom. By relaxing this configuartion the three adjacent atoms in (111) direction, i.e. the selfinterstitial, the Cd atom and the $nn$ host atom in (111) direction move more or less uniformly until the Cd atom settles in the bond-center position and the two host atoms about halfway between the interstitial and the substitutional positions. Therefore we refer to this as the Cd-split interstitial configuration, which is highly symmetric ($D_{3d}$ symmetry). It is shown in Fig.\ref{fig:6} (upper right), together with the exact positions of the neighboring atoms (Fig.\ref{fig:6} (bottom)).

\begin{figure}[t]
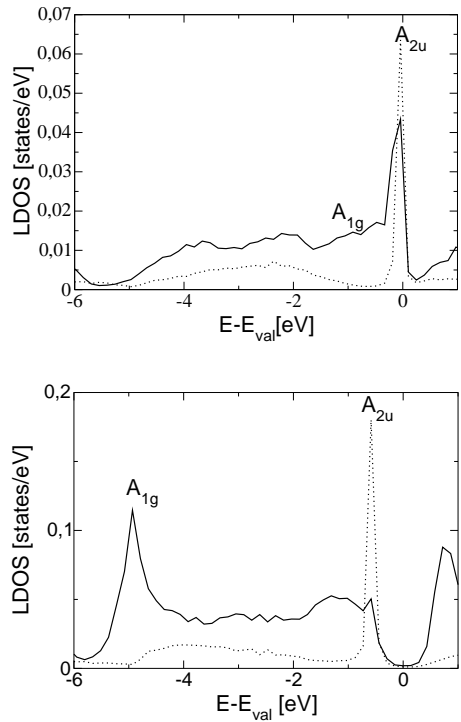

\center{\includegraphics*[scale=0.25]{vcsitea1_bw.eps}\\[0.5cm]
\includegraphics*[scale=0.25]{dos_cd_in_ge_zga_a1-anteil_bw.eps}
\caption{
\label{fig:7} LDOS in the bond-center cell for the cases that it is unoccupied by the Cd impurity (upper figure, a) and  that it is occupied by Cd (lower figure, b). The position of all other atoms are the same in both figures as given for the split configuration in Fig. 6 (upper right).}}
\end{figure}

The electronic structure of this complex can again be understood by considering the Cd atom as a Cd$^{2+}$ ion. Fig.\ref{fig:7} shows the LDOS in the cell of the Cd-atom for this configuration, in Fig.\ref{fig:7}a without the presence of the Cd atom and in Fig.\ref{fig:7}b in the presence of the Cd. Only the LDOS in the $A_{1g}$ and $A_{2u}$ sub-spaces are shown, since the local $E_u$ and $E_g$ contributions are small and not of relevance for the EFG. These states are mostly localized at the two interstitials and the 6 neighboring host atoms. 

\begin{figure}[hb]
\center{
\includegraphics*[scale=0.30]{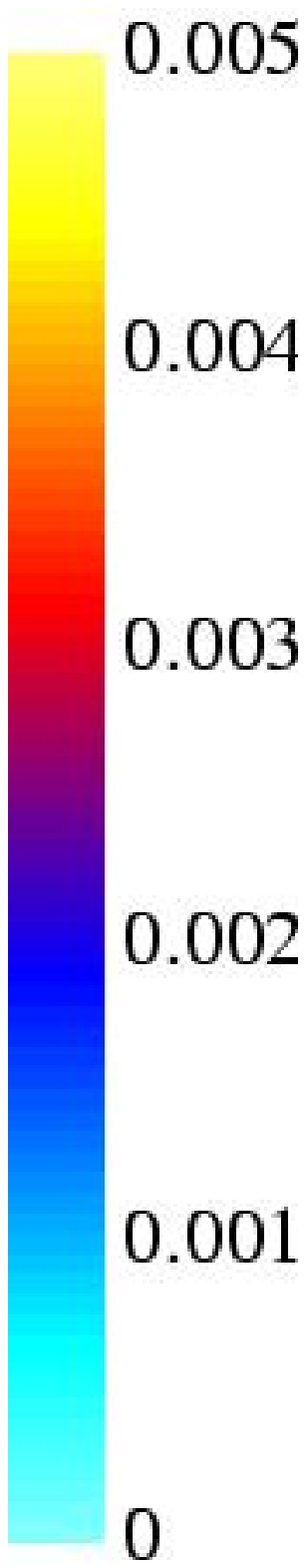}
\includegraphics*[scale=0.52]{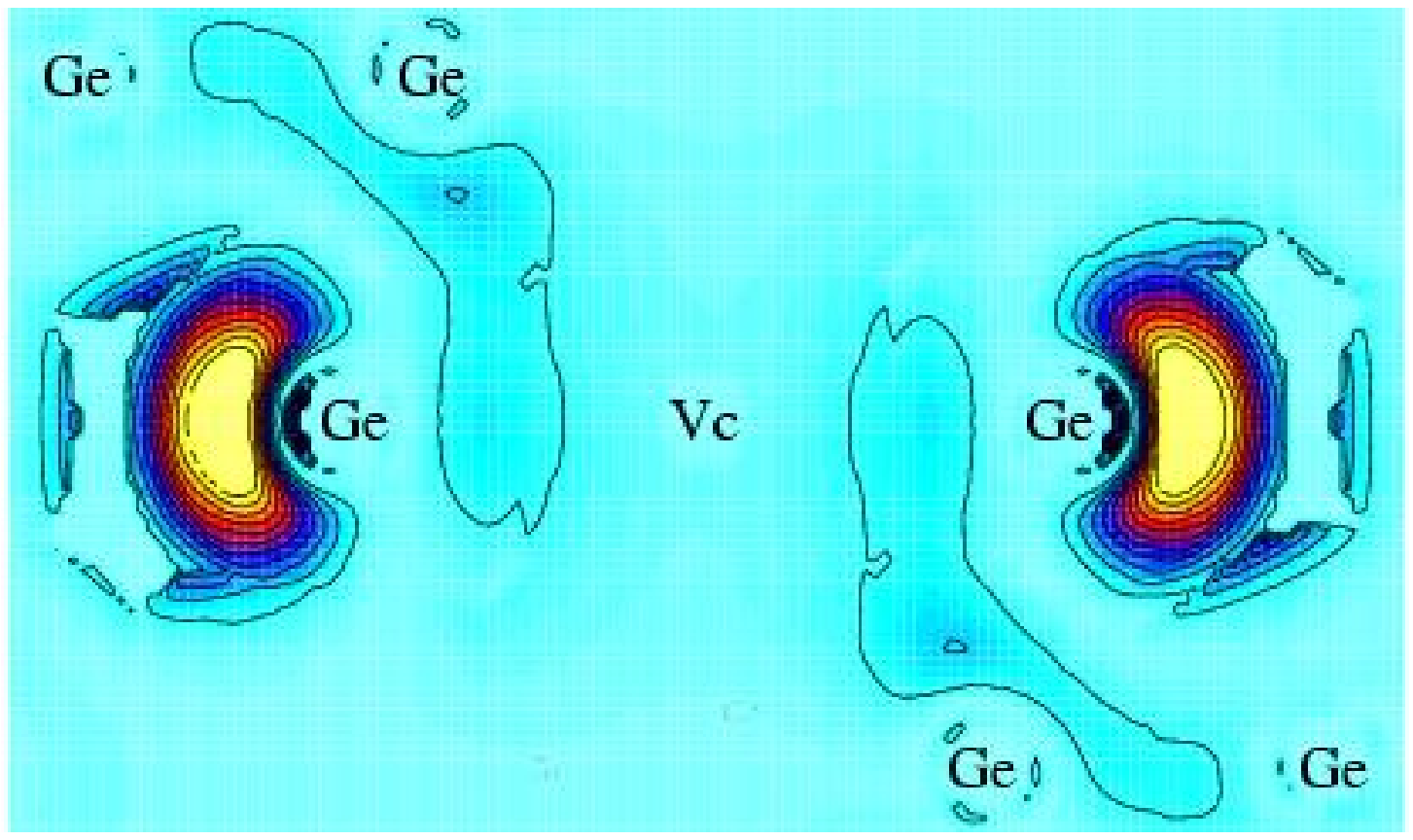}\\
\includegraphics*[scale=0.30]{farbsk.ps}
\includegraphics*[scale=0.52]{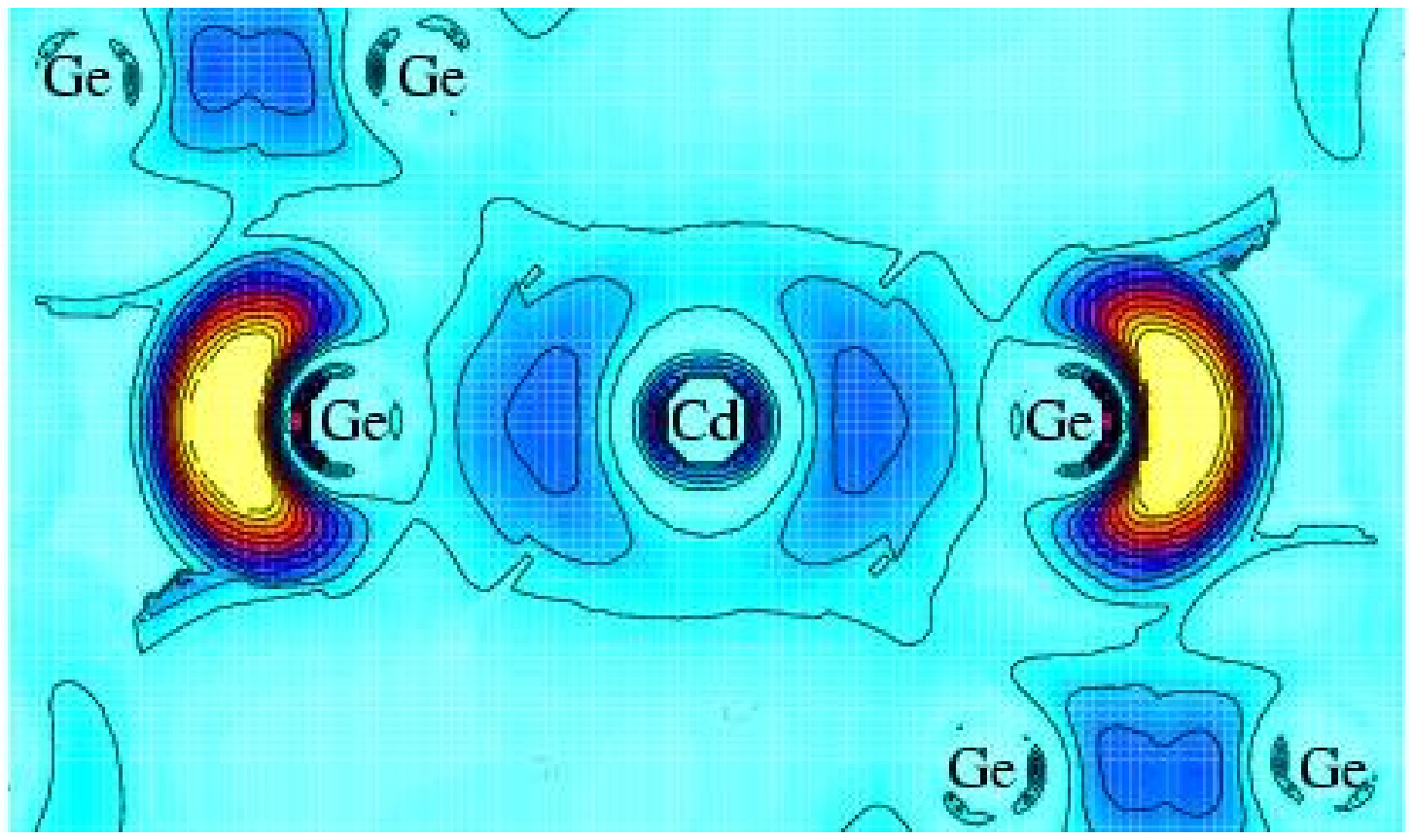}\\
\caption{
\label{fig:8} (color-online) A contour plot in the xz-plane of the $A_{2u}$ state close to the Fermi energy (see Fig.\ref{fig:7}) for the Cd split interstitial complex: Fig. 8a (upper panel) in the absence of the Cd impurity, Fig. 8b (lower panel) in the presence of the Cd on the bond-center position.}}
\end{figure}

For the two shifted host atoms without Cd, we see in Fig.\ref{fig:7}a two sharp peaks of $A_{1g}$ and $A_{2u}$ symmetry at $E_F$. These are symmetric and antisymmetric combinations of $p_z$--dangling bond states of the two split atoms, pointing into the interstitial region. A contour plot in the x-z plane of the DOS of these states is shown in Fig.\ref{fig:8}a. Since the two dangling bond states do practically not overlap, the $A_{1g}$ and $A_{2u}$ states are nearly degenerate. By inserting the Cd$^{2+}$ ion, the bonding $A_{1g}$ state is shifted to lower energies and becomes more localized at the Cd site, while the $A_u$ states is only slightly shifted to lower energies (Fig.\ref{fig:7}b). Both the $A_{1g}$ and $A_{2u}$ states are fully occupied yielding a neutral configuration. The contour plot of the $A$-state close to $E_F$, being dominated by the $A_{2u}$-state, shows Cd-$p_z$ orbitals forming a bridge between the two dangling bond states, with a nodal plane perpendicular to the $z$-axis (Fig.\ref{fig:8}b).

The calculated EFG of the Cd-split interstitial complex is quite large, i.e. -395 MHz in Ge and -345 MHz in Si. The strong anisotropy arises from the two split interstitial atoms. 
The major contribution to the EFG arises from the $p_z$ orbital in the $A_{2u}$-state, shown in the contour plot of Fig.\ref{fig:8}b; in addition small contributions come from $p_x$ and $p_y$ orbitals belonging to $E_u$ states.
The $A_{1g}$ state gives nearly no contribution, since the dominant $s$-state is isotropic and the $d_{z^2}$-admixture is small.
The calculated EFG of $-395$ MHz for the Cd-spit interstitial in Ge is in good agreement with the observed frequency\cite{Siele:98.2,Feuser:90.1} $\nu_2 = \pm 415$ MHz, which has been assigned by Haesslein et al.\cite{Siele:98.2} to a Cd-interstitial, but by Feuser et al.\cite{Feuser:90.1} to a Cd-vacancy complex. The calculated value of $- 345$ MHz for the interstitial complex in Si is more difficult to compare with the experimental data, which have been reviewed by Wichert et al.\cite{Wichert:89.3}. From our calculations we can exclude the axial symmetric and $< 111 >$ oriented complex $De3$ with $\pm 448$ MHz; moreover we can also exclude the defects $De 2$ since it is not axial symmetric, and  $De 1$ since it refers to the Cd-split vacancy complex. A good match is only obtained with the unidentified defect $X 2$ with $\pm 334$ MHz, being axial symmetric and $< 111 >$-oriented. 

\section{SUMMARY}

In this paper we have investigated the electronic and geometrical structure of complexes of the radioactive probe atom Cd with intrinsic defects in Si and Ge. In particular we have calculated the electric field gradient (EFG) of the Cd probe atom in these complexes, with the aim of assigning measured EFG frequencies to calculated defect complexes. The calculations are based on density functional theory in the local density approximation and apply the full-potential KKR Green's function method as well as the pseudo potential plane wave method. Both methods give nearly identical results for the atomic configurations. 

The most simple complex, a substitutional Cd-atom adjacent to a vacancy on a nearest neighbor site, is found to be unstable and to relax into a Cd-split vacancy configuration: the Cd-atom sits on the bond centered site, with the vacancy split into two half-vacancies on the neighboring substitutional sites. As seen from the Cd atom, this complex is very isotropic and results in a rather small EFG. The values for the single negatively charged complex, i.e. $- 55.7$ MHz in Ge and $-28$ MHz in Si, are in excellent agreement with measured EFGs ($\pm 54$ MHz in Ge and $\pm 28$ MHz in Si) and also the charge state agrees with conclusions of Haesslein et al.\cite{Siele:98.2}. This close agreement confirms the existence of the calculated Cd-split vacancy configuration with very small EFG. By investigating the most simple complex of a Cd-atom with an interstitial, i.e. a substitutional Cd atom with a selfinterstitial on the adjacent tetrahedral site, we find that this complex is also unstable and decays spontaneously into a Cd atom on the bond-center position and the two host atoms shifted from the substitutional site halfway to the tetrahedral position, thus forming a Cd-split interstitial complex aligned in $< 111 >$-direction. For Ge the calculated EFGs of $- 415$ MHz agrees well with the experimental value of $\pm 395$ MHz assigned by Haesslein et al. to a Cd-interstitial complex. The experimental situation in Si is not so clear.

These unusual split configurations are typical for intrinsic complexes with oversized impurities, as we will demonstrate in a future paper for a series of vacancy complexes with oversized and ``normal'' impurities. 

\begin{acknowledgments}
We thank Th. Wichert and R. Sielemann for helpful and motivating discussions, R. Sielemann in particular for suggesting the Cd-split vacancy configuration to us.
 \end{acknowledgments}


\begin{thebibliography}{99}
\bibitem{Siele:98.2}
Helmut Haesslein, Rainer Sielemann and Christian Zistl, Phys. Rev. Lett. {\bf 80 \rm}, 2626, (1998).
\bibitem{Feuser:90.1}
U. Feuser, R. Vianden and A.F. Pasquevich, Hyperfine Interactions {\bf 60 \rm}, 829, (1990).
\bibitem{Wichert:89.1}
Th. Wichert and M.L. Swanson, J. Appl. Phys. {\bf 66 \rm},3026, (1989).
\bibitem{Wichert:89.2}
Th. Wichert, M. Gr\"ubel, G. Keller, N. Schulz and H. Skudlik, Appl. Phys. A {\bf 648 \rm}, 59, (1989).
\bibitem{Wichert:99.1}
Th. Wichert, in Semiconductors and Semimetals, Academic Press, Volume {\bf 51B \rm}, Chapter 6, (1999).
\bibitem{Forkel:92.1}
D. Forkel, N. Achtziger, A. Baurichter, M. Deicher, S. Deubler, M. Puschmann, H. Wolf, W. Witthuhn,  Nucl. Instr. and Meth. in Phys Res. B {\bf 63 \rm}, 217, (1992).
\bibitem{Acht:93.1}
N. Achtziger and W. Witthuhn, Physical Review B {\bf 47 \rm}, 6990, (1993).
\bibitem{Akai:90.1}
Hisazumi Akai, Masako Akai, S. Bl\"ugel, B. Drittler, H. Ebert, Kiyoyuki Terakura, R. Zeller and P.H. Dederichs, Progress of Theoretical Physics, Supplement 101, 11, (1990). 
\bibitem{Forkel:87.1}
D. Forkel, F. Meyer, W. Witthuhn, H. Wolf, M. Deicher and M. Uhrmacher, Hyperfine Interactions {\bf 35 \rm}, 715, (1987). 
\bibitem{Deicher:87.1}
M. Deicher, G. Gr\"ubel, E. Recknagel, H. Skudlik and Th. Wichert, Hyperfine Interactions {\bf 35 \rm}, 719, (1987).
\bibitem{Wichert:89.3}
Th. Wichert, M. Deicher, G. Gr\"ubel, R. Keller, N. Schulz and H. Skudlik, Applied Physics A {\bf 48 \rm}, 59, (1989).
\bibitem{Siele:98.1}
R. Sielemann, Nucl. Instr. and Meth. in Phys Res. B {\bf 146 \rm}, 329, (1998).  
\bibitem{Siele:97.1}
Ch. Zistl, R. Sielemann, H. Haesslein, S. Gall, D. Br\"aunig and J. Bollmann, Material Science Forum {\bf 258-263 \rm}, 53, (1997).
\bibitem{Watkins:75.1}
G. D. Watkins, Physical Review B {\bf 12 \rm}, 4383, (1975).
\bibitem{Vosko:80.1}
S. H. Vosko, L. Wilk and M. Nusair, Can. J. Phys. {\bf 58 \rm}, 1200, (1980). 
\bibitem{Korr:47.1}
J. Korringa, Physica {\bf 13 \rm}, 392, (1947).
\bibitem{Kohn:54.1}
W. Kohn and N. Rostoker, Phys. Rev. {\bf 94 \rm}, 1111, (1954).
\bibitem{Brasp:84.1}
R. Zeller and P.J. Braspenning, Solid State Communications {\bf 42 \rm}, No. 10, 701, (1982).
P.J. Braspenning, R. Zeller, A. Lodder and P.H. Dederichs, Physical Review B {\bf 29 \rm}, 703, (1984). 
\bibitem{Blaha:88.1}
P. Blaha, K. Schwarz and P.H. Dederichs, Physical Review B {\bf 37\rm}, 2792, (1988).
\bibitem{Blaha:88.2}
P. Blaha, K. Schwarz and P.H. Dederichs, Physical Review B {\bf 38\rm}, 9368, (1988).
\bibitem{kromen:01}
Winfried Kromen, Ph.D.-Thesis, RWTH Aachen (2001); published as Berichte des Forschungszentrums J\"ulich, Vol. 3887, ISSN 0944-2952 (2001)
\bibitem{K-B:82}
L.\ Kleinman, D.\ M.\ Bylander, Phys.\ Rev.\ Lett.\ {\bf 48}, 1425 (1982).
\bibitem{Bloechl:94}
P.~E.~Bl\"ochl, Physical Review B {\bf 50\rm}, 17953, (1994).
\bibitem{Monkhort-Pack:77} 
J.~D.~Pack and H.~J.~Monkhorst, Physical Review B {\bf 16\rm}, 1748, (1977).
\bibitem{Nikos:97.1}
N. Papanikolaou, R. Zeller, P. H. Dederichs and N. Stefanou, Physical Review B {\bf 55\rm}, 4157, (1997).
\bibitem{Settels:99.3}
M. Asato, A. Settels, T. Hoshino, T. Asada, S. Bl\"ugel, R. Zeller and P.H. Dederichs, Physical Review B {\bf 60}, 5202, (1999).
\bibitem{Zeller:95.1}
R. Zeller, P.H. Dederichs, B. Ujfalussy, L. Szunyogh and P. Weinberger, Physical Review B {\bf 52}, 8807, (1995).  
\bibitem{Zeller:97.1}
Rudolf Zeller, Physical Review B {\bf 55}, 9400, (1997).
\end{thebibliography}
\end{document}